\documentclass{article}

\usepackage{arxiv}

\usepackage[utf8]{inputenc} 
\usepackage[T1]{fontenc}    
\usepackage{hyperref}       
\usepackage{url}            
\usepackage{booktabs}       
\usepackage{amsfonts}       
\usepackage{nicefrac}       
\usepackage{microtype}      
\usepackage{graphicx}
\usepackage{natbib}
\usepackage{doi}

\title{Motion Illusion-like Patterns Extracted from Photo and Art Images Using Predictive Deep Neural Networks}

\author{ {Taisuke Kobayashi}\\
	Laboratory of Neurophysiology\\
	National Institute for Basic Biology\\
	\texttt{taisukekbys@gmail.com} \\
	\And
	{Akiyoshi Kitaoka} \\
	College of Comprehensive Psychology\\
	Ritsumeikan University\\
	\And
	{Manabu Kosaka} \\
	Code\_monsters group\\
	Laboratory of Neurophysiology\\
	National Institute for Basic Biology\\
	\And
	{Kenta Tanaka} \\
	Code\_monsters group\\
	Laboratory of Neurophysiology\\
	National Institute for Basic Biology\\
	\And
	{Eiji Watanabe} \\
	Laboratory of Neurophysiology\\
	National Institute for Basic Biology,\\
	Department of Physiological Sciences\\
	The Graduate University for Advanced Studies (SOKENDAI)\\
	\texttt{eiji@nibb.ac.jp, eijwat@gmail.com} \\
}

\renewcommand{\shorttitle}{}

\begin{document}
\maketitle

\begin{abstract}
In our previous study, we successfully reproduced the illusory motion of the rotating snakes illusion using deep neural networks incorporating predictive coding theory. In the present study, we further examined the properties of the networks using a set of 1500 images, including ordinary static images of paintings and photographs and images of various types of motion illusions. Results showed that the networks clearly classified illusory images and others and reproduced illusory motions against various types of illusions similar to human perception. Notably, the networks occasionally detected anomalous motion vectors, even in ordinally static images where humans were unable to perceive any illusory motion. Additionally, illusion-like designs with repeating patterns were generated using areas where anomalous vectors were detected, and psychophysical experiments were conducted, in which illusory motion perception in the generated designs was detected. The observed inaccuracy of the networks will provide useful information for further understanding information processing associated with human vision.
\end{abstract}

\section{Introduction}
Motion illusion is among the most impressive visual illusions\citep{Kitaoka:2017}. In motion illusions, motion is perceived even when the relative positions of the observer and the observed object are unchanged. The Fraser–Wilcox illusion (FWI; reported in 1979) is a representative example of motion illusion\citep{Fraser:1979} and comprises a spiral pattern design of repeating luminance gradients. The direction of illusory motion varies from person to person. Since their first identification, numerous FWI variations have been reported \citep[e.g.][]{Faubert:1999, Naor-Raz:2000, Kitaoka:2003}, and similar to the original design of FWI, they are composed of a basic structure of light and dark gradients, although with greater individual differences in the strength of perceived motion than in the direction of motion.

The mechanism by which illusory motion occurs has long been debated. One possibility is that contrast intensity \citep{Faubert:1999, Naor-Raz:2000, Conway:2005, Backus:2005} affects the processing speed of neurons and is converted into motion perception. Additionally, the effects of eye movements \citep{Murakami:2006} have been discussed, with no clear conclusion reached. Because illusory motion reportedly causes activity in some areas of the cerebrum involved in visual perception \citep{Conway:2005, Ashida:2012, Kuriki:2008}, cerebral involvement in the mechanism associated with illusory motion has been suggested. Moreover, behavioral experiments suggest the existence of perception of motion illusions in animals such as rhesus monkeys \citep{Agrillo:2015}, cats \citep{Baath:2014}, lions\citep{Regaiolli:2019}, guppies, zebrafish\citep{Gori:2014}, and fruit flies\citep{Agrochao:2020}. Therefore, studying the motion-perception mechanisms common to animals with developed visual systems has attracted increasing attention.

With the recent development of deep neural networks (DNNs), research on their use as a tool to study brain function has become more active \citep{Richards:2019}. Comparisons of findings related to DNNs with the operating principles of the brain represent an ability to analogize psychophysical or physiological processes using computational principles. Given that one of the original motivations for DNN research was an attempt to find the essence of the brain by artificially reproducing the functions of the nervous system, such analogizing continues to be relevant. DNNs have recently been applied in visual-perception research \citep{Funke:2021, Kriegeskorte:2015}. Additionally, in illusion research, illusion-like phenomena have been reported in DNNs for the flash-lag effect (a type of subjective contour) and color illusions \citep{Lotter:2020, Gomez-Villa:2019, Gomez-Villa:2020}. DNNs allow changes to the structure of a given network and the weights of the connections, which represent alterations that cannot be applied to a living brain.

Our research group has studied motion illusion by attempting to reproduce illusions using DNNs and comparing them with human perception. We previously focused on the relationship between the occurrence of an illusion and the predictive function of the brain \citep{Watanabe:2010, Watanabe:2018}. In that study, we constructed a DNN model incorporating predictive coding theory \citep{Lotter:2017} as a theoretical model of the cerebrum \citep{Kawato:1993, Rao:1999, Friston:2005} and trained by first-person-viewed videos \citep{Watanabe:2018}. The DNN model predicted motion in the rotating  snakes illusion to a degree similar to that of human perception, suggesting that the DNN model could be used as a tool for studying the subjective perception of motion illusion.

In the previous paper, we analyzed only the rotating snakes illusion as a representative example of motion illusion. In the present study, we analyzed a variety of motion illusions using the DNN model and attempted to generate predictive images using ordinary static image datasets that included photographs and paintings. Additionally, we conducted psychophysical tests on human subjects and compared predictions by the DNN model with the results from human perceptions.

\section{Methods}

\subsection{Deep Neural Networks}

The connection-weight model of a trained DNN (PredNet; written in Chainer) used in this study was identical to a 500K model described previously \citep{Watanabe:2018}. The model was obtained by training using 500,000 video frames.

\subsection{Test images}

To test the prediction of the DNN model, we prepared five groups of test image stimuli: motion illusions (n = 300), modern art paintings (n = 300), classic art paintings (n = 300), movable objects of photo pictures (n = 300), and still objects of photo pictures (n = 300). The motion illusions were originally generated by Drs. Akiyoshi Kitaoka \citep{Kitaoka:website} (299 images) and Eiji Watanabe \citep{Watanabe:2019} (1 image). The images of art paintings were randomly collected from wikiart (\url{https://www.wikiart.org}) according to their classification. The images of photo pictures were collected at random by icrawler \citep{Chen:2017}, which is a framework of web crawlers (license = ``noncommercial, modify,'' and keywords = ``car,'' ``building,'' ``cat,'' etc.), followed by manual classification as ``Movable objects'' (animals, vehicles, etc.) or ``Still objects'' (buildings, mountains, etc.). Images were trimmed and scaled down, and the final size of all images was adjusted to 160 × 120 pixels (width × height) to adapt the training images. The five groups of test image stimuli (1500 images in total) were shared as a ``Visual Illusions Dataset'' \citep{KobayashiWatanabe:2019}.

\subsection{Prediction}

The DNN model predicted the 22nd image (P1 image) with reference to 21 consecutive images, which were 21 images copied from one test image. The network then predicted the 23rd image (P2 image) with reference to 22 consecutive images, using the P1 image as the 22nd image. The optical flow vectors between the P1 and P2 images were then calculated using a customized Python program. Optical flaw analyses were performed using the Lucas–Kanade \citep{Lucas:1981} and Farneback \citep{Farneback:2003} methods (sparse and dense optical flow analyses, respectively) (Figure 4).

\subsection{Psychophysical experiment}

The visual stimuli used in the psychophysical experiment were created by removing them from a photo picture (image A) and a painting (image B) images (Figure 4). The number of repetitions in the units placed on the circle was set to 24, which was the same number of stimuli used in the psychophysical experiment in \citep{Hisakata:2008} (Figure 7, left).

The psychophysical experiment was designed based on the method of Hisakata et al. \citep{Hisakata:2008} and conducted using a program written in Python using OpenGL (v.3.1.5; \url{https://pypi.org/project/PyOpenGL/}). The subjects were asked to answer whether they saw the stimuli rotated clockwise (CW) or counterclockwise (CCW) by keyboard input using a two-alternative forced choice. The face of the subject was fixed at 50 cm from the screen, and only the right eye was used for viewing. A gazing point with a viewing angle of 1$^{\circ}$ was established at the center of a white background, and a stimulus with an outer diameter of 7$^{\circ}$ and an inner diameter of 1$^{\circ}$ was presented at 12$^{\circ}$ to the left of the center for 0.5 seconds. The subject looked at the gazing point and viewed the stimulus with their peripheral vision. When they responded, the next stimulus was played, but the stimulus was designed so that there was a minimum of 1 second between the presentation of the previous stimulus and the playback of the next stimulus.

To quantitatively examine the illusory motion of the stimuli, we intentionally rotated the stimuli and determined the conditions under which the illusory motion did not occur. We used two types of stimuli: the original image and its left–right reversed version. This was done to counteract the perceived rotation-velocity bias. To statistically analyze the responses, we presented the same condition multiple times with randomly varying stimulus types and rotation velocities. From the statistical data obtained by this procedure, we calculated the rotational velocity of the stimulus based on the same probability of receiving an answer that the stimulus was rotating in the CW and CCW directions.

Figure 7 shows the raw data of the psychophysical experiment. The horizontal axis represents the velocity at which the stimulus was intentionally rotated (with CCW as the positive direction), and the vertical axis represents the probability of responding that each stimulus was rotated CCW. The obtained psychometric curves were fitted using a cumulative Gaussian function to calculate the rotational velocity and the rotation-cancellation velocity when the probability was 0.5. The rotation-cancellation velocity is the velocity required to cancel the rotation of the presented image, and the direction of rotation due to the motion illusion of the image is the velocity multiplied by a minus. Therefore, we used the original and reversed stimuli to calculate the rotational velocity of the stimulus as follows:

\begin{equation}
\frac{1}{2}\{(\rm{static\; rotational\; velocity\; of\; the\; reversed\; stimulus}) - (static\; rotational\; velocity\; of\; the\; original\; stimulus)\}
\end{equation}

The subjects were the authors T.K. and E.W. plus one naïve subject (n = 3; all healthy subjects with normal vision). Informed consent was obtained from all participants. The experiment was conducted according to the rules established by the Ethics Committee of the National Institute for Physiological Sciences (permit No. 20A063). The intentional stimulus rotational velocities were set to a range of $-$2.1–$+$2.1$^{\circ}$/s, and the velocity intervals were 0.3$^{\circ}$/s. There were 30 presentations of stimuli under each condition and 900 presentations of a single stimulus under all conditions.

\subsection{Open-source software}
All program codes (DNN, optical flow analysis, and psychophysical stimulus presentation software), trained models, and stimulus images were released as open-source software at the following website.

\begin{enumerate}
\item DNN: \url{https://doi.org/10.6084/m9.figshare.5483710}
\item Optical flow analysis: \url{https://doi.org/10.6084/m9.figshare.5483716}
\item Psychophysics: \url{https://github.com/taikob/Motion_Illusion_test}
\item Trained model: \url{https://doi.org/10.6084/m9.figshare.11931222}
\item Stimulus images: \url{https://doi.org/10.6084/m9.figshare.9878663}
\end{enumerate}

\section{Results}
Figure 1 shows the examples of optical flow vectors detected in the images predicted by the model against the five stimulus groups using the Lucas–Kanade method. Although relatively large and/or well-aligned optical flow vectors were detected in the predicted images against motion illusions, relatively small optical flows were detected in images predicted against other groups. The direction of the motion vector detected from the motion illusions agreed with the direction of the illusory motion perceived by humans. Notably, the directed optical flows were detected not only in the illusion of many colors, shapes, and gradients but also in the illusion of simple white triangles (Figure 1, upper left). As a fundamental property of the methodology, the Lucas–Kanade method extracts objects with a characteristic shape from images as feature points and exploits them as the starting points of the optical flow.  Therefore, it was not very meaningful that the eyes and hands were selected, given that they were extracted as feature points and used as the starting point for the optical flow (e.g., Mona Lisa and President Obama).

For quantitative analysis, the frequency rates of the absolute values of the optical flow vectors detected from each image group were evaluated (Figure 2) and averages of the absolute values of the optical flow vectors for each image group were generated (Figure 3). The results were as follows: motion illusions, 0.71 ± 0.18 (arbitrary units; Lucas–Kanade) and 0.64 ± 0.034 (Farneback); modern art paintings, 0.052 ± 0.0029 and 0.24 ± 0.021; realistic art paintings, 0.036 ± 0.00088 and 0.092 ± 0.0014; movable object photographs, 0.035 ± 0.0013 and 0.11 ± 0.0020; and still object photographs, 0.037 ± 0.0013 and 0.11 ± 0.0026. These results indicate that larger optical flow vectors were detected in the motion illusion group relative to the other groups. This tendency did not change according to the use of either Lucas–Kanade or Farneback analyses. These findings suggested that the DNN model accurately classified illusory and other images.

However, as shown in Figure 2, relatively large optical flow vectors were also detected in images from groups other than that including motion illusion, although the number of examples was small. To investigate the cause of such exceptionally large optical flow vectors, two images (one photograph and one painting), in which notably large optical flows were predicted, were identified, and the P1 and P2 images were compared in detail. Figure 4 shows that two large optical flows were detected in the area of the building on the left side of the picture using the Lucas–Kanade method, whereas the Farneback method detected a dense optical flow in the same region. Similarly, for the painting, characteristic optical flows were detected on the columns on the left side of the painting (Figure 4). Figure 5 shows the distributions of the absolute values of the optical flow vectors for each image of the photograph and the painting. Most of the optical flows detected in images A and B were far from the frequency rate peaks of the optical flows detected in the photograph (still objects) and painting (realistic arts) image groups (Figure 2). The maximum absolute values of the optical flow vectors detected in each image and calculated by each analytical method were 0.68 and 4.0 (Lucas–Kanade and Farneback, respectively) and 0.32 and 1.5, respectively. Figure 6 shows a plot of the brightness values, where an exceptionally large optical flow was detected. Comparing P1 (Figure 6, green line) with P2 (Figure 6, blue line) revealed a shift in the patterns of the two brightness distributions. These results indicate that the DNN model incorrectly predicted motion for static images that a human would not recognize as moving.

We then hypothesized that the patterns of exceptionally large optical flows detected in photographs and paintings might exhibit characteristics of motion illusions. We focused the analysis on most of the motion illusions having a repeating structure. Therefore, the areas where the large optical flows were detected (Figure 6, boxed regions) were excised and reassembled into circular repeating structures (Figure 7), followed by the psychophysical experiments using three human subjects. After testing the effect of the reconstructed design on human perception, we found that these artificially created designs rendered a type of motion illusion (Figure 7). The strength of the detected rotational velocity differed among the three subjects (TK: A, $-$0.36 $^{\circ}$/s and B, 0.31 $^{\circ}$/s; TS: A, $-$0.48 $^{\circ}$/s and B, 0.96 $^{\circ}$/s; and EW: A, $-$0.18 $^{\circ}$/s and B, 0.29 $^{\circ}$/s). However, for the same design, there was no individual difference in the direction of the detected rotational velocity.

\section{Discussion}

In this study, we showed that the DNN model distinguished between an illusion and other ordinary photographs and paintings, although it occasionally predicted motion in some parts of the ordinary images (Figure 4). Interestingly, we were able to create new motion illusions from the target portions of these images. These were among the first illusions to be discovered with the aid of artificial intelligence (another example is a series of illusory designs generated by the evolutionary illusion generator \citep{Sinapayen:2021}). It is possible that there existed small unit structures in the motion illusions and that a background involving normal scenery might suppress the occurrence of illusory motion when the unit structures exist alone. We previously suggested the existence of unit structures in our recent study on the rotating snakes illusion and the Fraser–Wilcox illusion \citep{KobayashiWatanabe:2021}.

Many motion illusions present a “repetition” of unit structures. As noted, we presume that the presence of even one of these unit structures can potentially cause the perception of motion. However, no single unit structure alone can cause the perception of illusory motion, which suggests that local information might lead to the perception of motion only when it is combined with global information. Supporting evidence suggests that a wide range of brain regions, from V1 to MT+, are involved in the perception of motion illusions \citep{Ashida:2012}, with higher brain regions (e.g., MT+) thought to integrate information from a broader perspective than V1. The DNN model was capable of detecting motion flow in the unit structure embedded in photographs and paintings that was not perceived by humans, which could indicate an underdeveloped ability to integrate global information, such as that in higher brain regions. For artificial perception to be useful for basic research, further studies are required.

\bibliographystyle{unsrtnat}


\section*{Acknowledgments}
Special thanks to the psychophysics subject for the kindness and cooperation.

\section*{Author contributions statement}
T.K. and E.W. conceived the research and designed the project. M.K. and K.T. wrote Python programs. A.K. provided the illusory illustrations and inspired the project members. T.K. ran programs, conducted all the preparation for psychological experiments, and performed data analysis. E.W. wrote the manuscript. All authors reviewed the manuscript.

\section*{Funding}
This work was supported in part by a MEXT/JSPS KAKENHI Grant-in-Aid for Scientific Research.

\clearpage

\begin{figure}[ht]
\centering
\includegraphics{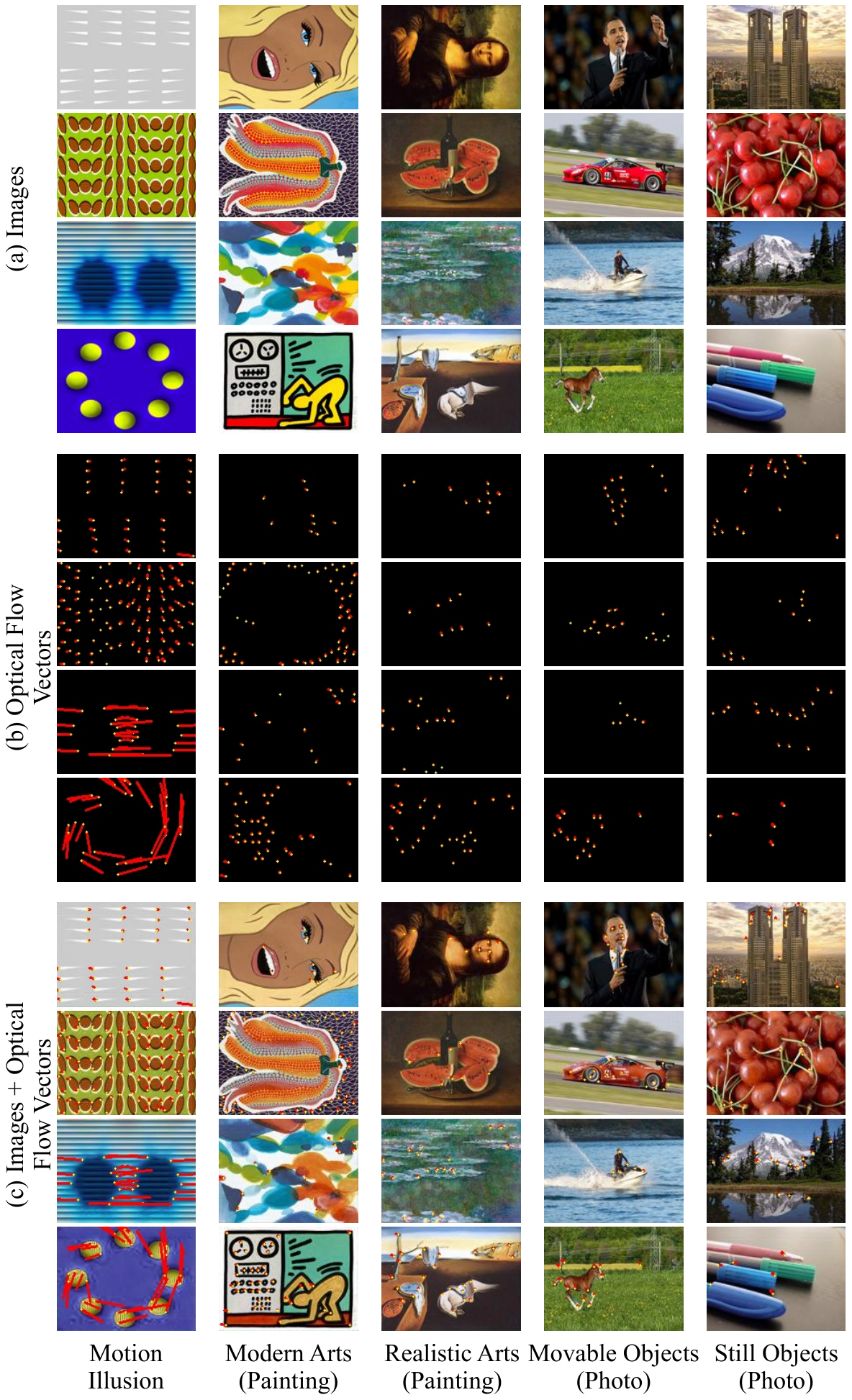}
\caption{Example images from the dataset and detected optical flows calculated using the Lucas–Kanade method. Four images are shown for each image group: motion illusion, modern arts (painting), realistic arts (painting), movable objects (photograph), and still objects (photograph). Each image group comprises a dataset of 300 images. (a) Original images input to the DNN model [image size: 160 × 120 pixels (width × height)]. (b) Detected optical flow vectors in the predicted images. Yellow points represent the starting points of the detected optical flow vectors, and red lines represent their directions and sizes. The length of the red line is 50 times that of the absolute value of the motion vector calculated using the Lucas–Kanade method. (c) The vectors were drawn over the predicted images.}
\end{figure}

\clearpage

\begin{figure}[ht]
\centering
\includegraphics{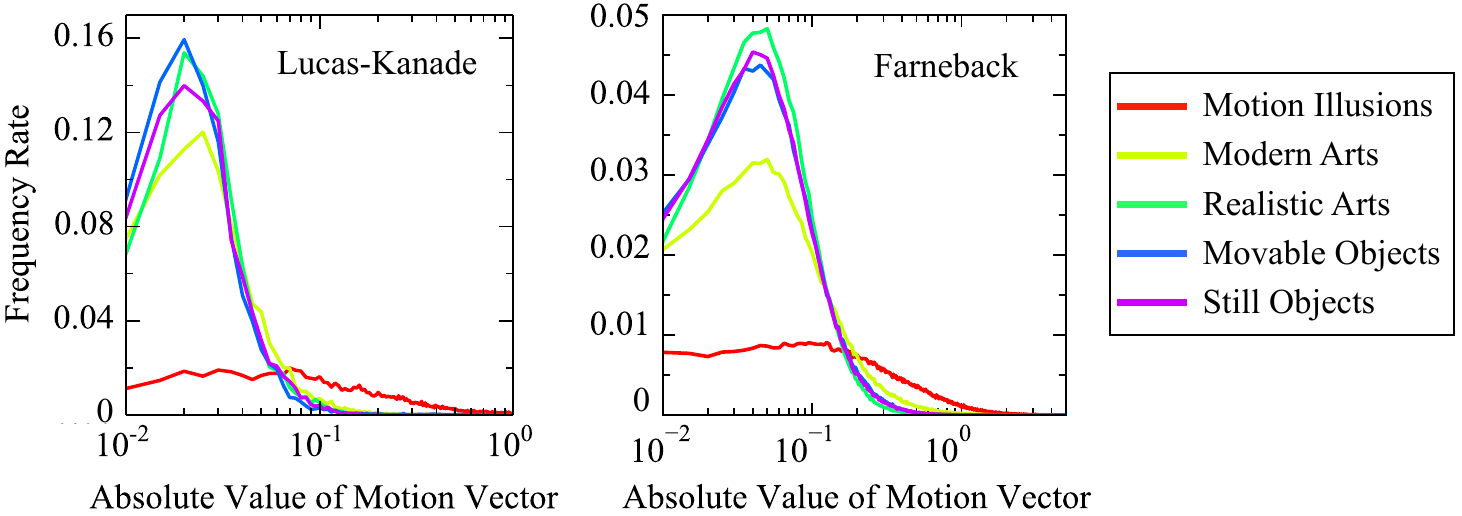}
\caption{Frequency rates of absolute values of the optical flow vectors for each image group. The vectors were calculated using the Lucas–Kanade (left) and Farneback (right) methods. The total number of vectors in each group was standardized to 1 for the frequency rate, and the size of the sampling window for the frequency rate was 0.01.}
\end{figure}

\clearpage

\begin{figure}[ht]
\centering
\includegraphics[scale=0.94]{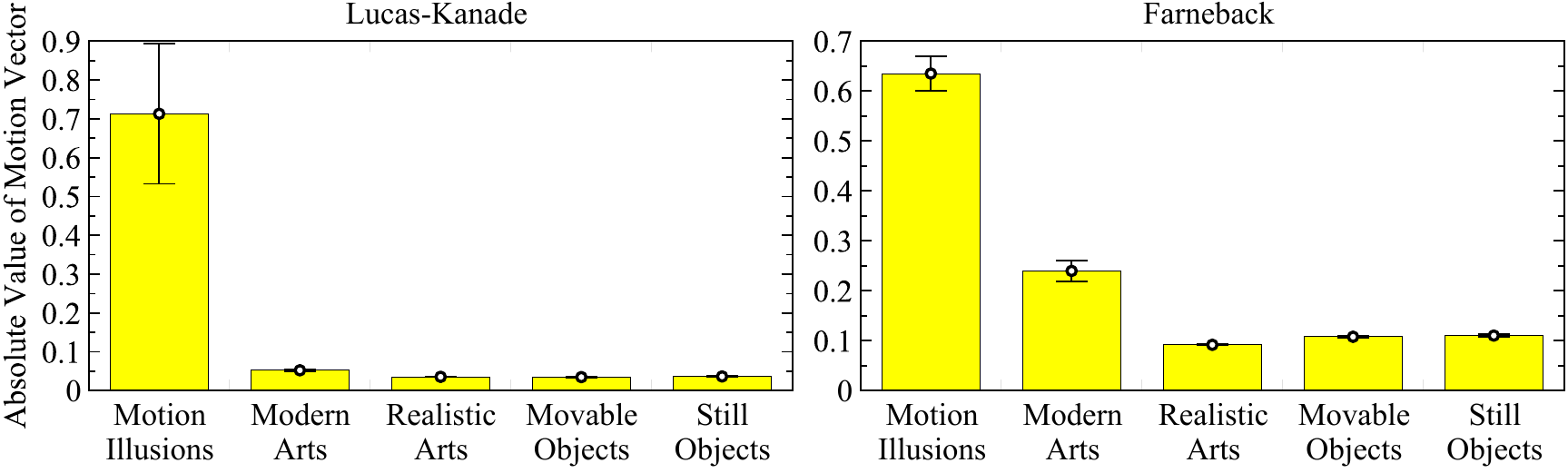}
\caption{Average values of the optical flow vectors for each image group. The vectors were calculated using the Lucas–Kanade (left) and Farneback (right) methods. The mean value of the vectors detected in each image was calculated, and the average values and standard errors of 300 images were calculated for each group.}
\end{figure}

\clearpage

\begin{figure}[ht]
\centering
\includegraphics[scale=0.95]{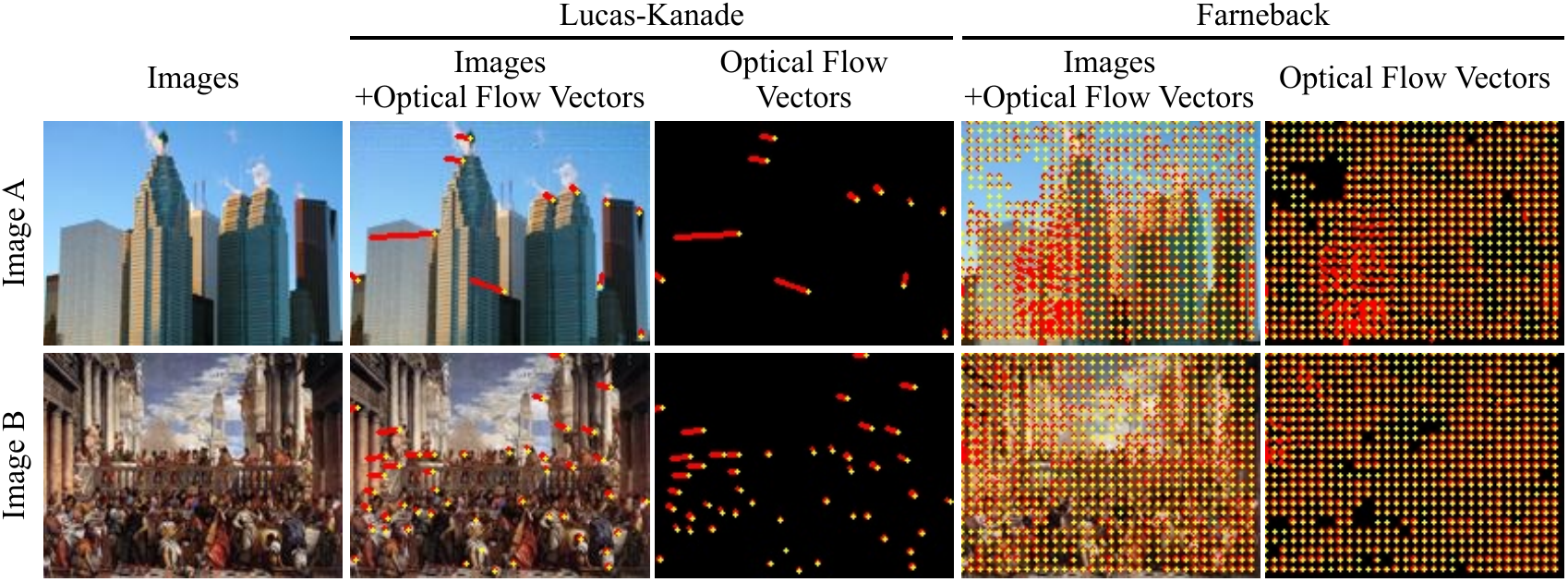}
\caption{Two examples of exceptional image stimuli from which the DNN model detected relatively large optical flow vectors. Image A was derived from photograph images and image B from painting images. Yellow points represent the starting points of the detected optical flow vectors, and red lines represent their mean directions and sizes. The length of the red lines is 50 and 4 times that of the absolute values of the motion vectors calculated using the Lucas–Kanade and Farneback methods, respectively. The original images input to the DNN model are shown on the far left.}
\end{figure}

\clearpage

\begin{figure}[ht]
\centering
\includegraphics{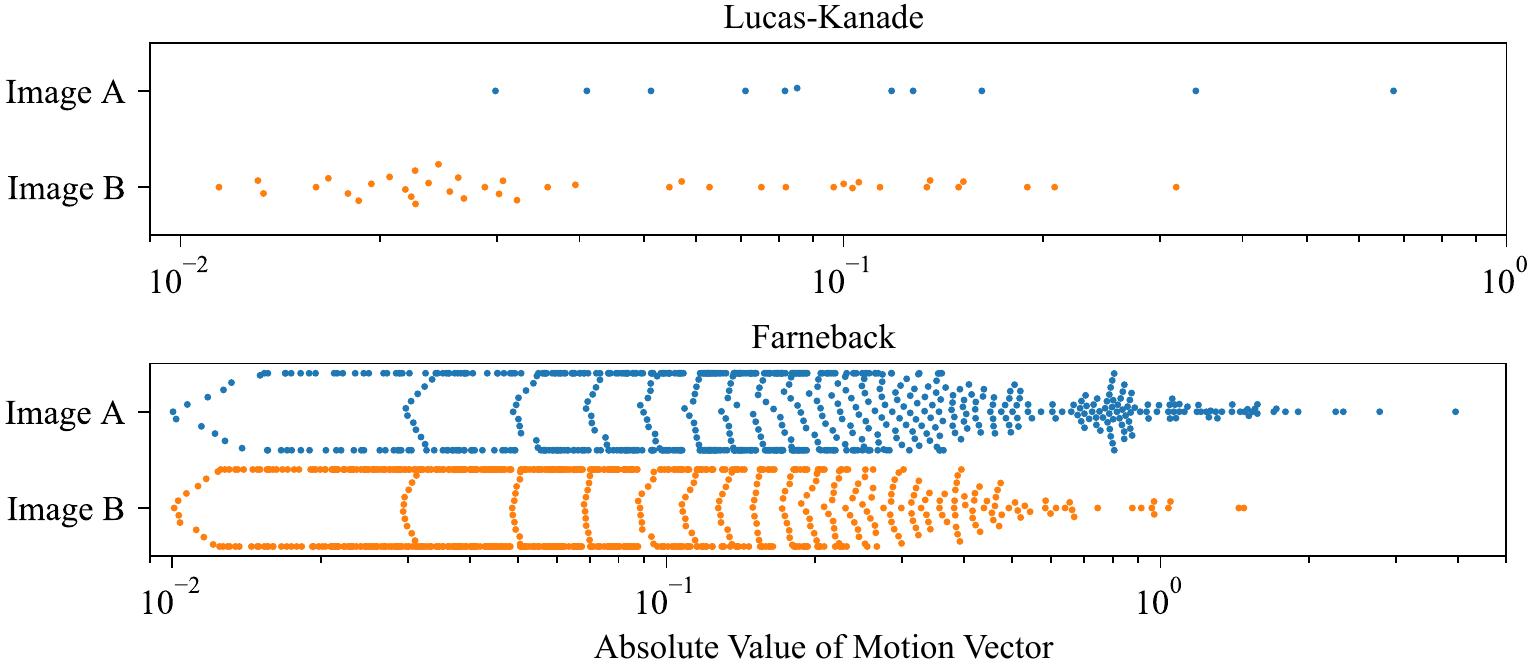}
\caption{Absolute value distribution of the optical flow vectors calculated from each image (A and B; as shown in Figure 4). To compare the distributions of all image-derived vectors (graphs in Figure 2), large vectors (rarely observed in the non-illusion group) were detected in the two images.}
\end{figure}

\clearpage

\begin{figure}[ht]
\centering
\includegraphics{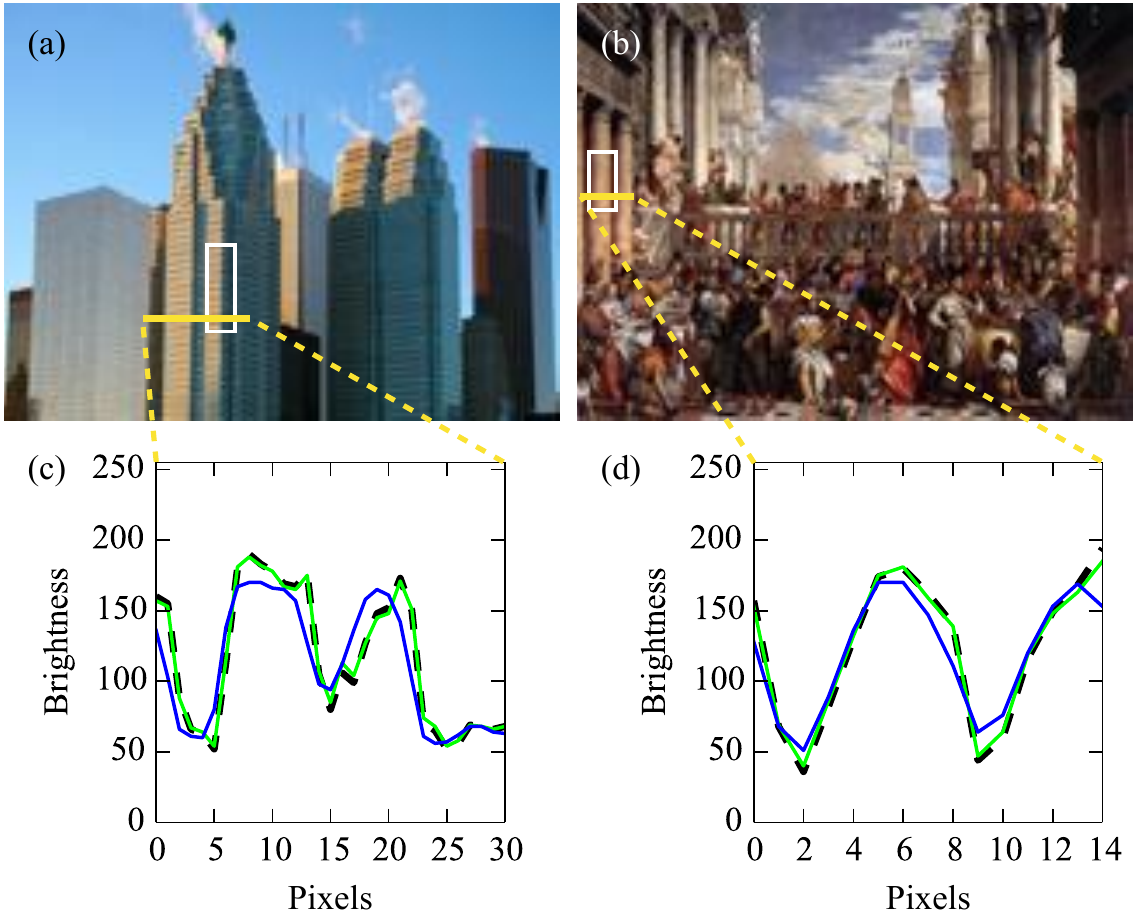}
\caption{Brightness distributions of images A and B. The predicted images (P1 and P2) were calculated using the DNN model following input of the original images. These images were converted into grayscale images using OpenCV and brightness values were calculated. Brightness values in images (a) A and (b) B and where large optical flows were detected (Figure 4) are plotted [(c) and (d), respectively] along yellow lines. The dotted line represents the original image, the green line represents the P1 image, and the blue line represents the P2 image. The box in each image was used to create motion-illusion-like designs.}
\end{figure}

\clearpage

\begin{figure}[ht]
\centering
\includegraphics{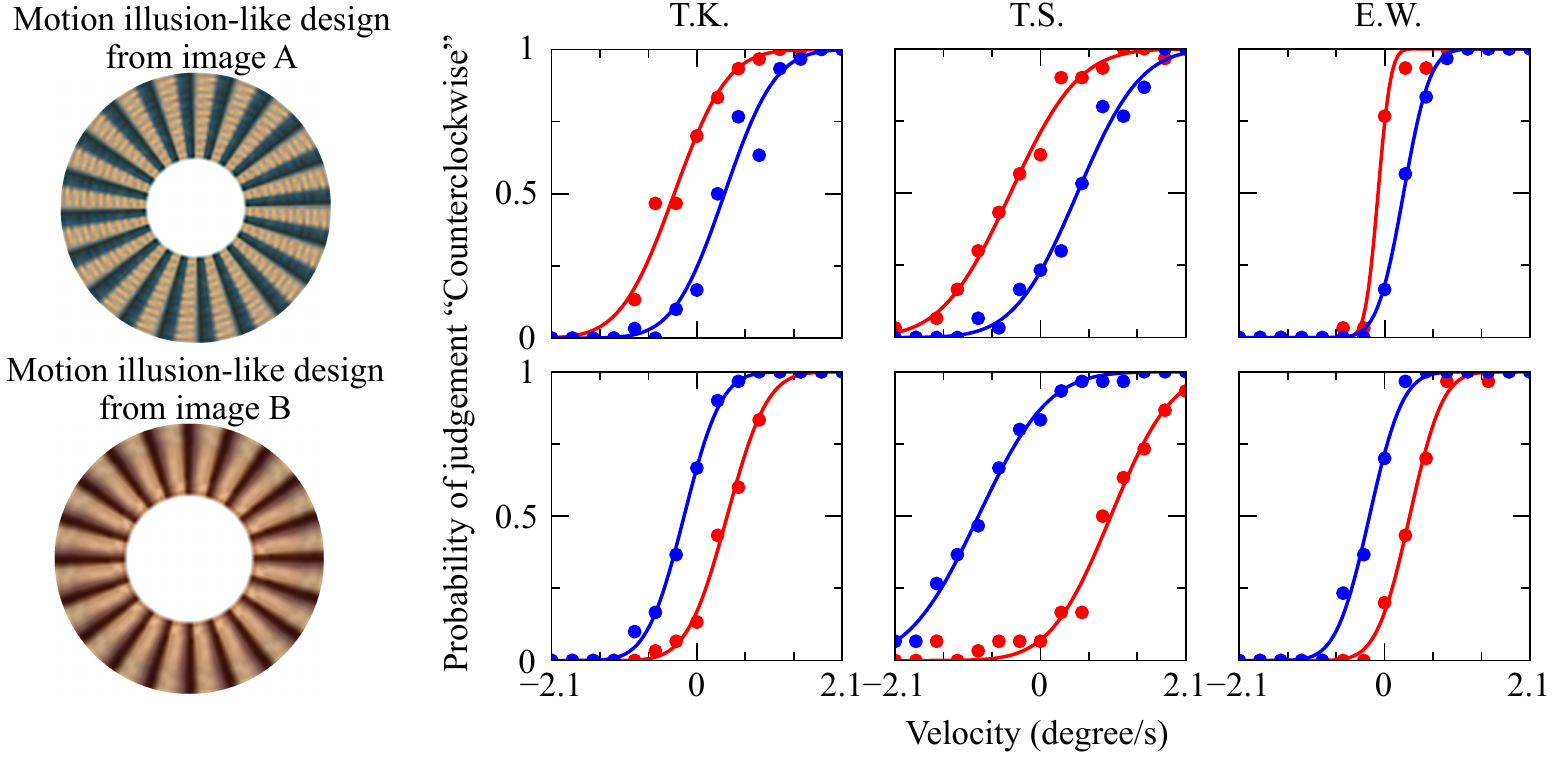}
\caption{The psychophysical experiments. The boxes in images A and B (Figure 6) were excised to create two motion-illusion-like designs (left), followed by psychophysical experiments using three subjects. Each psychometric curve was fitted with the cumulative Gaussian function using the least-squares method. The probability of seeing counterclockwise (CCW) rotation was plotted against the rotational velocity. Separate charts for each subject correspond to different eccentricities. Red and blue circles and the curves represent the raw probability data and best-fit curve for the CCW and clockwise (CW) stimulus, respectively. In the case of the design from image A, the subjects showed a high probability of answering that it was rotating CCW, whereas for image B, subjects showed a high probability of answering that it was rotating CW.}
\end{figure}

\end{document}